\documentclass[12pt]{iopart}
\usepackage{graphpap}
\usepackage[dvips]{graphicx}
\usepackage[dvips]{graphics}
\usepackage{color}

\begin{document}

\title{Coulomb oscillations in three-layer graphene nanostructures}
 \author{J. G\"uttinger, C. Stampfer, F. Molitor, D. Graf, T.~Ihn, and K.~Ensslin}
 
 \address{Solid State Physics Laboratory, ETH Zurich, 8093 Zurich, Switzerland}
 \date{ \today}
 \ead{guettinj@phys.ethz.ch}
 \begin{abstract}
We present transport measurements on a tunable three-layer graphene single electron transistor (SET). The device consists of an etched three-layer graphene flake with two narrow constrictions separating the island from source and drain contacts. Three lateral graphene gates are used to electrostatically tune the device. An individual three-layer graphene constriction has been investigated separately showing a transport gap near the charge neutrality point. The graphene tunneling barriers show a strongly nonmonotonic coupling as function of gate voltage indicating the presence of localized states in the constrictions.
We show Coulomb oscillations and Coulomb diamond measurements proving the functionality of the graphene SET. A charging energy of $\approx 0.6~$meV is extracted. 

 \end{abstract}

 \maketitle

\newpage
\section{Introduction}

Carbon materials, such as carbon nanotubes and graphene have attracted increasing interest in the past decades, which is mainly due to 
their unique electronic properties.
Strong suppression of electron backscattering in carbon nanotubes~\cite{ando98} and graphene~\cite{nov05,zha05} make both materials interesting for 
future high mobility nanoelectronic applications~\cite{gei07, Avouris07}. 
Their low atomic weight and the low nuclear spin concentration, arising from the $\approx 99$\% natural abundance of $^{12}$C are good premises for having weak spin orbit and hyperfine couplings. These make carbon nanomaterials also promising candidates for future spintronic devices~\cite{Awsch07,Tomb07} and spin-qubit based quantum computation~\cite{Loss98,Elzer04,Petta05,Trau07}.
It has been shown recently that nanotubes exhibit a topologically induced 
spin-orbit coupling~\cite{kue08}, which 
is directly related to their cylindrical shape~\cite{and00,hue06}.
In graphene and few-layer graphene such flux accumulating (circumferential) trajectories should not be present, leading to a legitimate hope for much weaker spin-orbit interaction and thus possible applications for spin-based quantum information processing.
However, graphene and few-layer graphene quantum devices 
are still in their infancy since 
it is hard to confine carriers in these semi-metallic materials using electrostatic potentials.

Here we report on Coulomb oscillations and Coulomb diamond
measurements on an etched and fully tunable three-layer graphene single electron transistor.
Single electron transistors (SETs) consist of a small island connected via tunneling barriers
to source and drain contacts~\cite{kou97}. First few-layer graphene SETs have been formed by Schottky barrier contacts on graphitic flakes~\cite{sco05} and just very recently etched single-layer graphene structures~\cite{stampfer08,pon08} have been fabricated to demonstrate Coulomb blockade. 

\section{Device and fabrication}
The investigated SET device is shown in Figs.~1(a)-(c) and consists of a three-layer graphene structure. Two $\approx60~$nm wide constrictions separate the graphene island from the source (S) and drain (D) contacts. The two constrictions are separated by about $1~\mu$m, while the area of the island is A$\approx0.1~\mu$m$^2$. Two lateral side gates SG1 and SG2 allow to change the three-layer graphene barriers electrostatically and independently. The potential on the island can be separately changed by an additional side gate denoted as plunger gate (PG). The highly doped Si substrate is used as a back gate (BG) giving control over the overall Fermi level. Note that on the same flake also an additional single constriction has been fabricated (see inset in Fig.~2a).

 \begin{figure}[hbt]\centering
\includegraphics[draft=false,keepaspectratio=true,clip,%
                   width=0.85\linewidth]%
                   {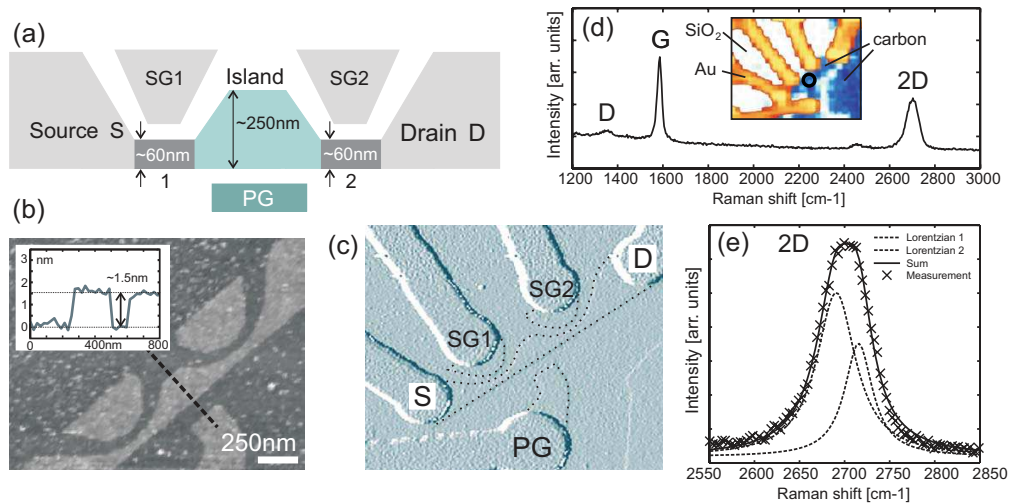}                  
\caption[FIG1]{(color online)
(a) Schematic illustration of the single electron transistor device. (b) SFM image of the device after reactive ion etching. The inset shows a cross section along the dashed line. (c) SFM image of the final device with metal contacts. The dotted lines indicate the circumference of the graphene structures. (d) Confocal Raman spectra recorded on the island after fabrication as highlighted in the Raman image shown as inset. The laser spot size is approx. 400~nm. (e) The 2D line can be approximated by two Lorentzians. For more details see text.} 
\label{trdansport}
\end{figure}

The devices have been fabricated by mechanical exfoliation of graphite flakes on Si substrates covered with 295~nm thick SiO$_2$ as described in Ref.~\cite{nov04}. The individual graphite flakes were patterned by electron-beam (e-beam) lithography using 90~nm PMMA as resist and a subsequent reactive ion etching step by an Ar/O$_2$ plasma (9:1). Fig.~1(b) shows a scanning force microscope (SFM) image of the etched three-layer graphene flake (bright area). A second e-beam lithography step followed by metalization and lift-off is used to place 2~nm Ti and 50~nm Au electrodes as shown in Fig.~1(c). 

Confocal Raman spectroscopy measurements~\cite{dav07a} have been used to determine the thickness of the graphitic flake, i.e. the number of graphene layers. A Raman spectrum recorded at the center of the three-layer graphene island is shown in Fig.~1(d). This spot is marked by a circle in the Raman image~\cite{sta07a} of the device which is shown as an inset in Fig.~1(d). In this Raman image, white areas are attributed to the silicon oxide, bright (yellow) areas to the contacts and dark (blue) areas to the three-layer graphene. Apart from the elevated background due to nearby metal contacts the spectrum shows pronounced G and 2D lines typical for sp$^2$ graphitic materials. The defect induced D line arises from the edges of the flake inside the area of the laser spot, which has a diameter of about $400~$nm. 
It is known from earlier experiments \cite{dav07a,fer06,gupta07} that the lineshape of the 2D peak and the intensity ratio of G/2D provides direct insight into the number of graphene layers of the investigated flake. Thus the 2D line is analyzed in more detail as shown in Fig.~1(e). According to Ref.~\cite{dav07a} we can fit the 2D peak either by 1, 2 or 4 Lorentzian(s), just depending on the number of graphene layers. It turns out that the measured spectra can be best fitted by the sum (straight line) of two Lorentzians [dashed lines in Fig.~1(e)]. The center of the two Lorentzians are offset by $\Delta \omega = 25.0\pm0.5~$cm$^{-1}$. In addition, the ratio between the integrated intensity of the G and the 2D line is $\approx 0.57$. Both measures, including the relative height of the two Lorentzians provide strong support for having three-layer graphene. We can exclude bilayer graphene, since for two layers the integrated intensity ratio is $\mathrm{G}/\mathrm{2D} = 0.38\pm0.02$~\cite{dav07a}, but there is still a small chance of having four layer graphene. For four-layer (six-layer) graphene the two Lorentzians are offset by $\Delta \omega = 26\pm2~$cm$^{-1}$ ($\Delta \omega = 28\pm2~$cm$^{-1}$) and the integrated intensity ratio is $\mathrm{G}/\mathrm{2D} = 0.63\pm0.08$ ($\mathrm{G}/\mathrm{2D} = 0.7\pm0.1$)~\cite{dav07a}. It is found that our mesurements fit best to the value of three layers where the following values have been reported $\Delta \omega = 25.4\pm1.5~$cm$^{-1}$ and $\mathrm{G}/\mathrm{2D} = 0.53\pm0.05$. This is also in good agreement with the step height of $\approx1.5~$nm taken from the SFM data [inset and dashed line in Fig.~1(b)]. However, four-layer graphene can not be completely excluded here. In the following we refer to our sample to consist of three layers of graphene. The overall conclusions would not change if the sample had indeed four layers of graphene.
 
  

If not stated explicitly, measurements were performed at 2~K in a variable temperature cryostat by applying a symmetric source-drain bias voltage $V_{\mathrm{bias}}$ (DC and small superimposed AC component) and measuring the source-drain current. The samples have been heated up in vacuum to $135^\circ$C for 12~h before cool down to eliminate undesired atoms on the sample surface as much as possible.

\section{Results and discussion}

 \begin{figure}[t]\centering
\includegraphics[draft=false,keepaspectratio=true,clip,%
                   width=0.9\linewidth]%
                   {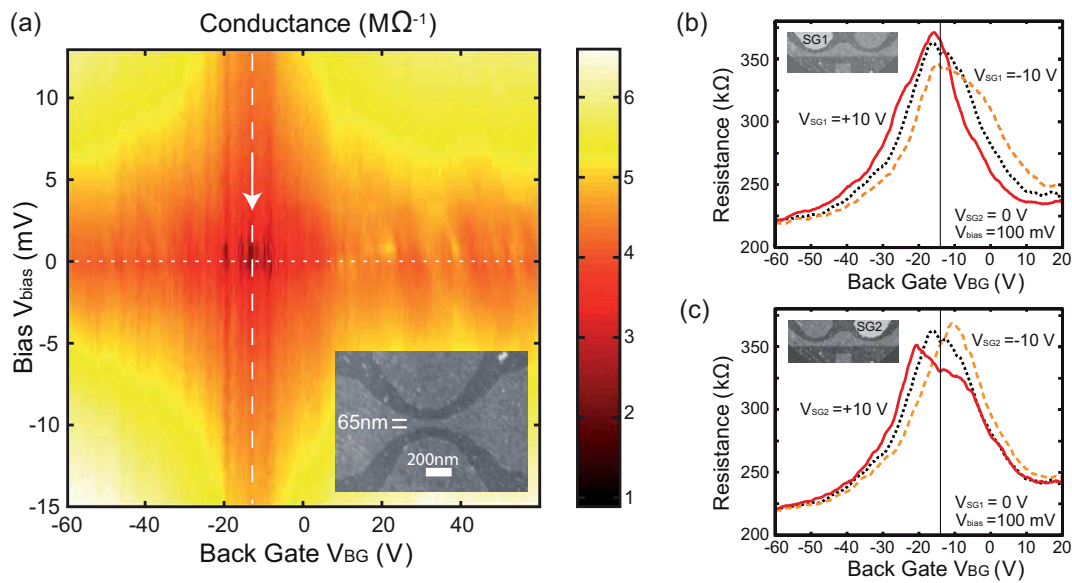}                  
\caption[FIG2]{(color online)
High bias source-drain current measurements. (a) Conductance as a function of bias and back gate voltage for an individual three-layer graphene constriction ($\approx 65$~nm wide). A SFM image of the structure is shown as inset. At low bias the opening of a transport gap is observed around $V_{\mathrm{BG}} \approx -17$~V.  (b) and (c) show back gate characteristics for different side gate voltage configurations. In (b) $V_{\mathrm{SG2}}=0$ and the influence of SG1 is investigated for $V_{\mathrm{SG1}} = 10~\textrm{V (line), }0~\textrm{V (dotted), and }-10~$V (dashed line). In (c) $V_{\mathrm{SG1}}=0$ and the influence of SG2 is investigated for $V_{\mathrm{SG2}} = 10~\textrm{V (line), }0~\textrm{V (dotted), and }-10~$V (dashed line). A difference of the local doping of the two constrictions can be observed. For more information see text.} 
\label{fig2}
\end{figure}
We first discuss measurements on a single constriction. This structure has been fabricated from the same graphene flake as the SET. The investigated constriction has a width of $\approx 65~$nm, as shown in the inset of Fig.~2(a). Differential conductance measurements as function of  $V_{\mathrm{bias}}$ and $V_{\mathrm{BG}}$ (i.e., Fermi level) reveal the presence of a transport gap around the charge neutrality point (at $V_{\mathrm{BG}} \approx -17$~V, see white arrow). This is indicated by (i) conductance suppressions near the charge neutrality point, including strong fluctuations for low bias, which might be due to resonances in the three-layer graphene constriction [see dark regions in Fig.~2(a)] and (ii) the significant conductance non-linearity as a function of increasing bias. This result is very similar to what has been measured in single-layer graphene nanoribbons~\cite{han07, chen07}, where it has been argued that a nanoribbon-width dependent effective energy gap (i.e. transport gap) dominates the transport near the charge neutrality point~\cite{sols07}. 
For the measured three-layer graphene constriction we observe the onset of a transport gap which is smaller than the thermal energy ($\approx k_{\mathrm{B}}T = 0.2$~meV), since we do not observe any gap induced pinch off at small bias voltages. The gap is therefore significantly smaller than the $\approx 4$~meV, which has been reported for 65~nm wide single-layer graphene constrictions~\cite{sols07,han07}. However, most importantly, three-layer graphene constrictions exhibit a transport gap which can be used to form tunneling barriers for defining a three-layer graphene island, very much like in single-layer graphene~\cite{stampfer08,pon08}.

Measurements on the SET are performed first in the high bias regime ($V_{\mathrm{bias}} = 100$~mV), where transport is not suppressed by the two constrictions. This allows us
to investigate different regimes in the back- and side gate parameter range.
Figs.~\ref{fig2}(b) and~2(c) show the (two-point) source-drain resistance under the influence of different side gate potentials [$V_{\mathrm{SG1}}$ is stepped in Fig.~2(b) and $V_{\mathrm{SG2}}$ in Fig.~2(c)]. The dotted traces show measurements where the side gate voltages have been set to zero. The dashed and the solid lines correspond to measurements where negative ($V_{\mathrm{SG1,2}}$ = -10~V) and positive ($V_{\mathrm{SG1,2}}$ = 10~V) side gate potentials have been applied respectively. The resistance shows a peak around $V_{\mathrm{BG}} = -15$~V, which we identify as the conductance minimum at the overall charge neutrality point of the significantly n-doped sample. For increasing or decreasing back gate voltage the resistance decreases. This 
can be well explained by the (linear) carrier density increase as function of the back gate voltage when moving away from the charge neutrality point~\cite{nov04}.
By applying different side gate potentials the peak height, width, and position change. 
For example, setting $V_{\mathrm{SG1}}$ = 10~V and sweeping $V_{\mathrm{BG}}$ the resistance peak becomes higher and narrower, while it is less pronounced and broader for $V_{\mathrm{SG1}}$ = -10~V [see Fig.~2(b)]. The opposite behavior is observed for applying $\pm$10~V to SG2, as shown in Fig.~2(c).
In both cases a positive side gate voltage leads to a down shift in back gate voltage, whereas a negative voltage leads to an up shift. 

These transport characteristics can be well explained by assuming (i) that the transport in this regime is dominated by the two constrictions~1 and~2 and (ii) that the two constrictions are differently doped, i.e., they exhibit two different (local) charge neutrality points. Here constriction~2 is slightly more n-doped than constriction~1.     
According to Ref.~\cite{molitor07} graphene side gates work well for locally changing the 
carrier density, i.e., for locally shifing the charge neutrality point. 
Moreover, we assume (will be shown below) that the crosstalk of SG1 and SG2 on constrictions~2 and~1 is negligible.
Therefore, a positive potential on SG1 reduces the
doping inbalance between constriction~1 and~2 [see Fig.~2(b)]. This increases the sample	homogeneity and results in a high and narrow resistance peak. In contrast, by applying $V_{\mathrm{SG1}}$ = -10~V the doping difference increases and the resistance peak significantly
shrinks and broadens.  
The measurements shown in Fig.~2(c) can be explained similarly. Additionally, we can now extract the relative lever ams of the side gates to the back gate with respect to the transport dominating constrictions given by $\Delta V_{\mathrm{SG1,2}}/\Delta V_{\mathrm{BG}} \approx $2.8.
From now on SG1 is operated in a positive and SG2 in a more negative voltage regime in order to match the doping levels in constrictions~1 and~2, which might lead to a more symmetric coupling of the island to source and drain. 

\begin{figure}[hbt]\centering
\includegraphics[draft=false,keepaspectratio=true,clip,%
                   width=0.75\linewidth]%
                   {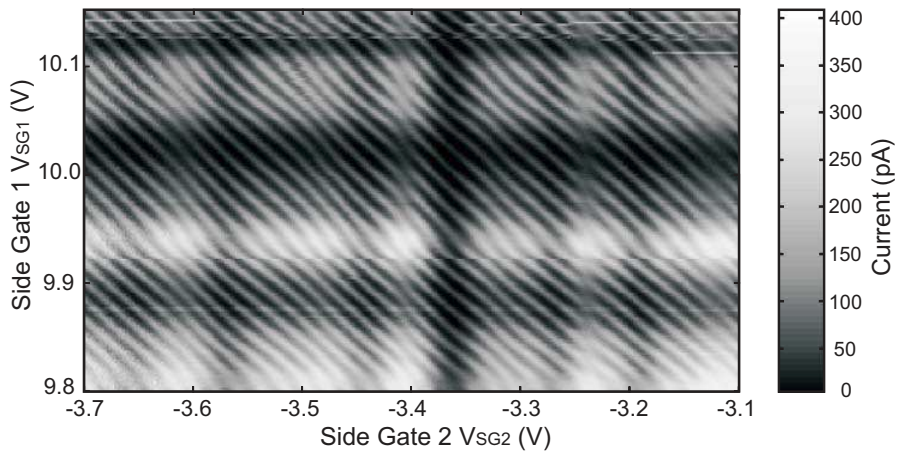}                  
\caption[FIG3]{
Source-drain current plotted as function of the two side gates voltages $V_{\mathrm{SG1}}$ and $V_{\mathrm{SG2}}$ for constant bias voltage ($\textrm{V}_{\mathrm{bias}} = 200~\mu\textrm{V}$). Measurements are taken at $\textrm{V}_{\mathrm{BG}} = -13.47~\mathrm{V}$ and $\mathrm{V}_{\mathrm{PG}} = 0~\mathrm{V}$.} 
\label{fig3} 
\end{figure}

We now discuss the low bias transport properties. 
Fig.~\ref{fig3} shows a measurement of the source-drain current as function of both side gates
$V_{\mathrm{SG1}}$ and $V_{\mathrm{SG2}}$ performed at $V_{\mathrm{bias}} = 200~\mu$V and $V_{\mathrm{BG}} = -13.47$~V. 
We observe sequences of horizontal and vertical stripes of suppressed current and current resonances.
Their direction in the $V_{\mathrm{SG1}}$-$V_{\mathrm{SG2}}$ plane indicates that their physical origin has to
be found within constriction~1 (vertical stripes) or constriction~2 (horizontal stripes).
The current exhibits even finer equidistantly spaced
resonances which are almost equally well tuned by both side gates. We therefore
attribute these resonances to states localized on the island between the barriers. It
will be shown below that these resonances are Coulomb oscillations due to charging of the three-layer graphene island.
The overall behavior is very similar to what has been observed in a single layer graphene SET~\cite{stampfer08}. 
 \begin{figure}[hbt]\centering
\includegraphics[draft=false,keepaspectratio=true,clip,%
                   width=0.85\linewidth]%
                   {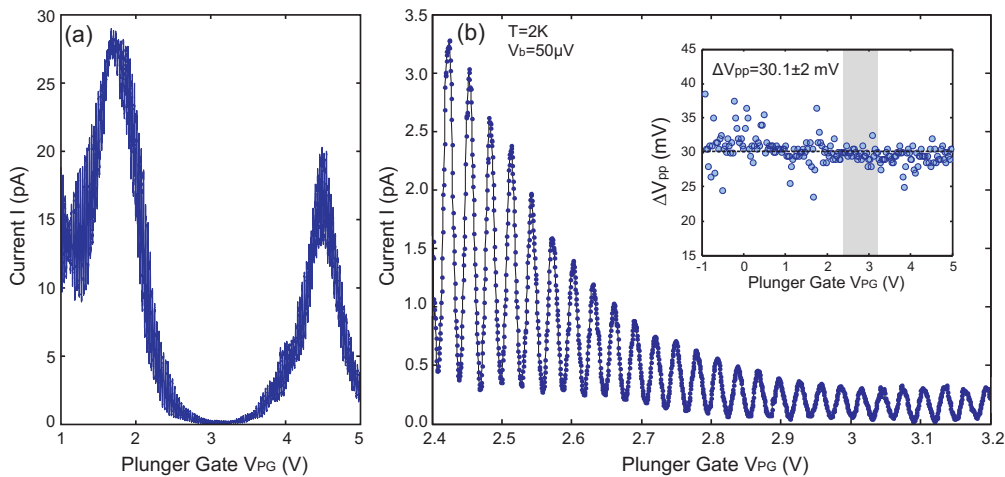}                  
\caption[FIG1]{(color online)
Coulomb oscillations [Fig.(b)] as a function of the plunger gate voltage $V_{\mathrm{PG}}$. Although the background is strongly modulated by constriction resonances [see Fig.~(a)] an almost constant peak spacing is observed over more than 150 oscillations [see inset in Fig.~(b)]. Measurements are taken at $V_{\mathrm{SG1}} = 6~\mathrm{V} - 0.27 V_{\mathrm{PG}}$, $V_{\mathrm{SG2}} = -2.37~$V, $V_{\mathrm{BG}} = -10.51~$V and $V_{\mathrm{bias}} = 50~\mu$V.} 
\label{fig4}
\end{figure}

Coulomb oscillations are further investigated by modulating the plunger gate voltage $V_{\mathrm{PG}}$ and simultaneously compensating its influence on the constrictions by the side gates respectively. Fig.~4 shows the current as a function of $V_{\mathrm{PG}}$. Here SG1 has been swept simultaneously following $V_{\mathrm{SG1}} = 6~\mathrm{V} - 0.27 V_{\mathrm{PG}}$. The plunger gate induced background modulation due to resonances in constriction~2 is negligible within this $V_\mathrm{PG}$ range and has therefore not been compensated. For these measurements $V_{\mathrm{bias}} = 50~\mu$V has been applied and the back gate voltage has been set to $V_{\mathrm{BG}} = -10.51~$V. Thus the Fermi energy in the source and drain contacts lies within the conduction band.  
In accordance with the measurement in Fig.~3 the current shows coarse and fine modulations. Again the larger oscillations with characteristic $V_{\mathrm{PG}}$ spacings of a few volts [see e.g. Fig.~4(a)] are attributed to transmission resonances in the constrictions while the fine current 
modulations at a voltage scale of around 30~mV are Coulomb oscillations, as shown in Fig.~4(b).
The elevated background at the left side in Fig.~4(b) is due to resonances in the constrictions. 
The inset in Fig.~4(a) shows the spacing of the Coulomb oscillations $\Delta V_{\mathrm{pp}}$ as function of the plunger gate voltage, which has been swept over more than 150 periods. The gray marked region corresponds to the oscillations shown in Fig.~4(b). The mean Coulomb peak spacing is  $\Delta V_{\mathrm{pp}} = 30.1\pm2~$mV and part of the observed broadening might be due to the underlying modulation of the transmission through the narrow constriction [see correlation between Fig.~4(a) and inset in Fig.~4(b)]. The essentially constant peak spacing indicates that the three-layer graphene transistor compared to the single layer graphene SET~\cite{stampfer08} behaves much more like a metallic SET~\cite{for00}. One also needs to take into account that the three-layer device investigated here is larger than the single layer device presented in Ref.~\cite{stampfer08}. Therefore the single-level spacing which could give rise to peak spacing fluctuations is also smaller. Nevertheless the data in the inset of Fig.~4(b) resembles pretty much observations on metallic SET with the additional feature of superposed constriction resonances. Like for the single-layer graphene SET, some peak spacing fluctuations can be seen in Fig.~3 indicating that weak inhomogeneities exist within the dot including its edges.
\begin{figure}[t]\centering
\includegraphics[draft=false,keepaspectratio=true,clip,%
                   width=0.75\linewidth]%
                   {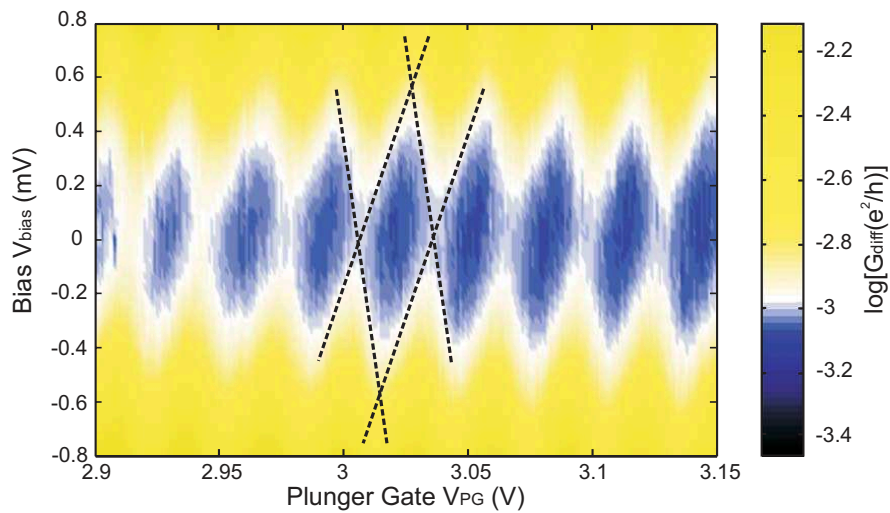}                  
\caption[FIG1]{(color online)
Coulomb diamond measurements. From the size of the Coulomb diamonds we estimate a value for the charging energy of
around 0.6~meV. The DC bias is modulated with a $50\mu$V AC component allowing for a direct lock-in amplifier measurement of the differential conductance. The measurement is performed with $V_{\mathrm{SG1}} = 6~\mathrm{V} - 0.27\cdot V_{\mathrm{PG}}$, $V_{\mathrm{SG2}} = -2.37~$V and $V_{\mathrm{BG}} = -10.51~$V.} 
\label{fig05}
\end{figure}

Corresponding Coulomb diamond measurements~\cite{kou97}, i.e., measurements of the differential conductance ($G_{diff}=dI/dV_\mathrm{bias}$) as a function of bias voltage $V_{\mathrm{bias}}$ and plunger gate voltage $V_{\mathrm{PG}}$ are shown in Fig.~5.
For this measurement the same gate voltage configuration as in Fig.~4 has been used and an AC modulation of $50~\mu$V has been superimposed on to the DC bias. The differential conductance is plotted logarithmically as function of the plunger gate and bias voltage. From this measurement the charging energy is estimated to be $E_{\mathrm{C}} \approx 0.6~$meV. This corresponds to a total capacitance of $C_{\Sigma} = e^2/E_{\mathrm{C}} \approx 271~$aF. 
The electrostatic coupling capacitances of the different gates to the island are $C_{\mathrm{PG}} \approx 7.1~$aF, $C_{\mathrm{SG1}} \approx 8.5~$aF, $C_{\mathrm{SG2}} \approx 7.1~$aF and $C_{\mathrm{BG}} \approx 28.45~$aF. The capacitance of the island to the back gate $C_{\mathrm{BG}}$ can be compared with a plate capacitor model leading to a capacitance $C=\epsilon_0\epsilon A/d \approx 12$~aF. The difference to the capacitance obtained from the measurement can be explained by edge effects which are not accounted for in the simple plate capacitor model. These measurements are in accordance to what has been observed for a single-layer graphene SET~\cite{stampfer08}. There the island area is significantly smaller, leading to a larger charging energy ($\approx$3.5~meV) and a larger discrepancy between the back gate capacitance and the plate capacitor model is observed due to the even more enhanced edge effects of the smaller island.

 \begin{figure}[t]\centering
\includegraphics[draft=false,keepaspectratio=true,clip,%
                   width=0.95\linewidth]%
                   {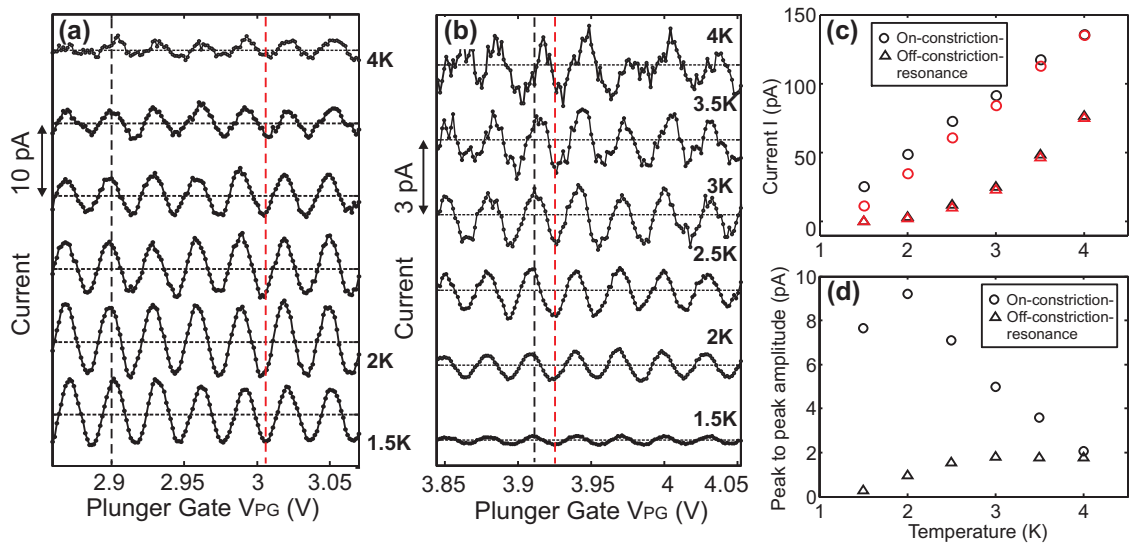}                  
\caption[FIG1]{(color online)
Temperature dependence of Coulomb oscillations (COs) in two different regimes. (a) COs sitting on top of a constriction resonance ("on-constriction-resonance"). (b) COs in a regime where the current is strongly supressed by the constrictions ("off-constriction-resonance"). (c) Current measured at the plunger gate voltages indicated by the dashed lines in (a) and (b) for both regimes. (d) Temperature dependence of the COs peak to peak amplitude. Measurements taken at $V_{\mathrm{SG1}} = 5.067~\mathrm{V} - 0.27 (V_{\mathrm{PG}}-3.437~\textrm{V})$, $V_{\mathrm{SG2}} = -1.588~$V, $V_{\mathrm{BG}} = -10.51~$V and $V_{\mathrm{bias}} = 200~\mu$V.  } 
\label{fig06}
\end{figure}

Finally, we investigate the temperature dependence of the Coulomb oscillations and the background (i.e., constriction) resonances in two different plunger gate regimes. Within the first regime the background is strongly elevated, whereas in the second regime the background is strongly suppressed since we are between two constriction resonances. The Coulomb oscillations on top of a constriction resonance are plotted for different temperatures in Fig.~6(a). Here the background has been subtracted and the traces are vertically offset by 10~pA for clarity. In Fig.~6(b) the data are presented in the same way for the "off-resonance" regime in the barrier transmission with a spacing of 3~pA between the traces. The change of the background current is shown in Fig.~6(c), where the current is plotted for two fixed PG voltages indicated by the vertical dashed lines in Fig.~6(a) and Fig.~6(b). On top of the constriction resonance (circles) the background current increases linearly with temperature, which does not hold for the regime with suppressed transmission ("off-constriction-resonance", triangles). There, the nonlinear current increase is attributed to raising and subsequent broadening of constriction resonances lifting the overall background current. Fig.~6(d) shows the averaged peak to peak Coulomb oscillation amplitudes. While the "on-constriction-resonance" amplitudes (circles) of the Coulomb oscillations are in general decreasing with increasing temperature the amplitudes of the "off-constriction-resonance" oscillations (triangles) are mainly limited by the transmission of the constrictions and therefore increase due to enhanced transmission (elevated background) with increasing temperature.

In conclusion, we have fabricated a tunable three-layer graphene single electron transistor based on an etched graphitic flake with lateral gates. Its functionality was demonstrated by observing clear and reproducible Coulomb oscillations. The tunneling barriers formed by three-layer graphene constrictions were investigated independently. From the Coulomb diamond measurements it was estimated that the charging energy of the three-layer graphene island is $\approx$0.6~meV, which is compatible with its lithographic dimensions. 
The overall behavior of the investigated device is very much like that observed for a single-layer graphene single electron transistor. The almost constant Coulomb peak spacing indicates the more metallic character of the three layer graphene SET.
These results open the way to more detailed studies of future graphene and few-layer graphene quantum devices.


{Acknowledgment ---}
The authors wish to thank R.~Leturcq, P.~Studerus, C.~Barengo and K.~Novoselov for
helpful discussions. Support by the ETH FIRST Lab and
financial support by the Swiss National Science Foundation
and NCCR nanoscience are gratefully acknowledged.



\section*{References}
\begin{thebibliography}{99}

\bibitem{ando98} 
Ando T, Nakanishi T and Saito R 1998 {\it J. Phys. Soc. Jpn.}~{\bf 67} 2857

\bibitem{nov05}
Novoselov K S, Geim A K, Morozov S V, Jiang D, Katsnelson M I, Grigorieva I V, Dubonos S V and Firsov A A 2005 {\it Nature}~{\bf 438} 197-200

\bibitem{zha05}
Zhang Y, Tan Y W, Stormer H L and Kim P 2005 {\it Nature}~{\bf 438} 201-4

\bibitem{gei07}
Geim A K and Novoselov K S 2007 {\it Nat. Mater.}~{\bf 6} 183

\bibitem{Avouris07}
Avouris P, Chen Z H, Perebeinos V 2007 {\it Nature Nanotechnology}~{\bf 2} 605

\bibitem{Awsch07}
Awschalom D D and Flatté M E 2007 {\it Nature Physics} {\bf 3} 153 

\bibitem{Tomb07}
Tombros N, Jozsa C, Popinciuc M, Jonkman H T and van Wees B J 2007 {\it Nature} {\bf 448} 571

\bibitem{Loss98}
Loss D and DiVincenzo D P 1998 {\it Phys. Rev. A.} {\bf 57} 120

\bibitem{Elzer04}
J. M. Elzerman J M, Hanson R, Willems van Beveren L H, Witkamp B, Vandersypen L M K and Kouwenhoven L P 2004 {\it Nature} {\bf 430} 431-5

\bibitem{Petta05}
Petta J R, Johnson A C, Taylor J M, Laird E A, Yacoby A, Lukin M D, Marcus C M, Hanson M P and Gossard A C 2005 {\it Science} {\bf 309} 2180-84

\bibitem{Trau07}
Trauzettel B, Bulaev D V, Loss D and Burkard G 2007 {\it Nature Phys.} {\bf 3} 192

\bibitem{kue08}
Kuemmeth F, Ilani S, Ralph D C and McEuen P L 2008 {\it Nature} {\bf 452} 448 

\bibitem{and00}
Ando T 2000 {\it J. Phys. Soc. Jpn.} {\bf 69} 1757-63

\bibitem{hue06}
Huertas-Hernando D, Guinea F and Brataas A 2006 {\it Phys. Rev. B} {\bf 74} 155426

\bibitem{kou97}
Kouwenhoven L P et al 1997 {\it Electron transport in quantum dots (Mesoscopic
Electron Transport} ed Sohn L L, Kouwenhoven L P and Sch\"on G (Dotrecht: NATO
Series, Kluwer)


\bibitem{sco05} 
Bunch J S, Yaish Y, Brink M, Bolotin K and McEuen P L 2005 {\it Nano Lett.} {\bf 5-2} 287-90

\bibitem{stampfer08}
Stampfer C, G\"uttinger J, Molitor F, Graf D, Ihn T and Ensslin K 2008 {\it Appl. Phys. Lett.} {\bf 92} 012102

\bibitem{pon08} 
Ponomarenko L A, Schedin F, Katsnelson M I, Yang R, Hill E H, Novoselov K S and Geim A K 2008 {\it Science} {\bf 320} 320-58

\bibitem{nov04}
Novoselov K S, Geim A K, Morozov S V, Jiang D, Katsnelson M I, Dubonos S V, Grigorieva I V and Firsov A A 2004 {\it Science} {\bf 306} 666

\bibitem{dav07a} Graf D, Molitor F, Ensslin K, Stampfer C, Jungen A, Hierold C and Wirtz L  2007 {\it Nano Lett.} {\bf 7} 238

\bibitem{sta07a}
Stampfer C, B\"urli A, Jungen A and Hierold C 2007 {\it Physica Status Solidi B} {\bf 244-11} 4341-45

\bibitem{fer06} Ferrari A C  {\it et al} 2006 {\it Phys. Rev. Lett.} {\bf 97} 187401 

\bibitem{gupta07}
Gupta A, Chen G, Joshi P, Tadigadapa S and Eklund P C 2006 {\it Nano Lett.} {\bf 6} 2667

\bibitem{han07}
Han M Y, \"Ozyilmaz B, Zhang Y and Kim P 2007 {\it Phys. Rev. Lett.} {\bf 98} 206805

\bibitem{chen07}
Chen Z, Lin Y M, Rooks M J and Avouris P 2007 {\it Physica E: Low-dimensional Systems and Nanostructures} {\bf 40-2} 228-32

\bibitem{sols07}
Sols F, Guinea F and Castro Neto A H 2007 {\it Phys. Rev. Lett.} {\bf 99} 166803


\bibitem{molitor07} Molitor F, G\"uttinger J, Stampfer C, Graf D, Ihn T and Ensslin K 2007 {\it Phys. Rev. B.} {\bf 76} 245426-5


\bibitem{for00}
Furlan M, Heinzel T, Jeanneret B, Lotkhov S V and Ensslin K 2000 {\it Europhys. Lett.} ~{\bf 49} 369



%




%
%
%
%
%
%
%
%
%
%
%
%
%
%
%
%
%
%
%
%
%
%
%
%
%
%
%
%
%
%
%
%
%









%
%
%
%
%
%
%
%
%
%
%
%
%
%
%
%
%
%
%
%
%
%
%
%
%



%
%
%


\end{thebibliography}
\end{document}